# Proposal of Vital Data Analysis Platform using Wearable Sensor

Yoji Yamato[a,*]

[a]Software Innovation Center, NTT Corporation, 3-9-11 Midori-cho, Musashino-shi, 180-8585 Japan

*Corresponding Author: yamato.yoji@lab.ntt.co.jp

## Abstract

In this paper, we propose a vital data analysis platform which resolves existing problems to utilize vital data for real-time actions. Recently, IoT technologies have been progressed but in the healthcare area, real-time actions based on analyzed vital data are not considered sufficiently yet. The causes are proper use of analyzing methods of stream / micro batch processing and network cost. To resolve existing problems, we propose our vital data analysis platform. Our platform collects vital data of Electrocardiograph and acceleration using an example of wearable vital sensor and analyzes them to extract posture, fatigue and relaxation in smart phones or cloud. Our platform can show analyzed dangerous posture or fatigue level change. We implemented the platform. And we are now preparing a field test.

**Keywords**: IoT, Wearable Sensor, Cloud Computing, Spark Streaming, Real-Time Processing,

## 1. Introduction

Recently, IoT (Internet of Things) technologies have been progressed. IoT is the technology to attach communication functions to physical things, connect things to networks, analyze things data to enable automatic control. IoT application areas are wide such as manufacturing, supply chain, maintenance which Industrie4.0 [1] and Industrial Internet [2] target and also health care, agriculture, energy.

To utilize IoT data, IoT platforms also appeared to develop and operate IoT applications effectively. AWS IoT [3] is a platform to analyze IoT data on a cloud by integrating several Amazon Web Services. For example, Amazon Kinesis collects and delivers IoT data by MQTT(MQ Telemetry Transport) protocol to a cloud and Amazon Machine Learning analyzes those data by machine learning algorithms. To integrate IoT data and other services, there are some service coordination technologies such as [4]-[8].

In manufacturing or maintenance area, there are applications of appropriate timing maintenance actions based on monitored business machine statuses (e.g., KOMTRAX [9]), but in the healthcare area, real-time actions based on analyzed vital data are not considered sufficiently yet. Of course, there are applications to show daily statistical information such as calorie consumption using wearable sensor data such as amount of movement acquired by list band sensor, however it has not been able to utilize vital data to real-time actions.

There are two main causes. The first is proper use of analyzing methods. To utilize vital data in real-time, not only batch processing but also stream processing for continuous data and micro batch processing for bulk data of short period are needed, but it is not considered to apply health care industry sufficiently. The second is network cost. Because vital data is continuously generated, bandwidth to transfer them to a cloud is large.

Based on these backgrounds, in this paper, we propose a vital data analysis platform which resolves existing two problems based on open source HortonWorks Data Platform (HDP) [10] architecture to utilize vital data. Our platform collects workers' vital data of Electrocardiograph and acceleration using wearable vital sensor (e.g., hitoe [11]) and analyzes them to extract posture, fatigue and relaxation in smart phones or cloud. Our platform can show analyzed dangerous posture or fatigue level change.

The rest of this paper is organized as follows. In Section 2, we review existing IoT technologies. In Section 3, we propose and design a vital data utilization platform which resolves existing problems. We compare related work in Section 4. And we summarize the paper in Section 5.

## 2. Overview of IoT processing technologies

Because IoT technologies include a lot of topics such as sensor, actuator, big data, platform, communication protocol and so on, this section only introduces existing platform technologies and wearable sensor for IoT vital data analyzing applications.

To utilize IoT data collected by sensing technologies, AWS IoT [3] is a major platform. Amazon IoT can integrate each service of Amazon Web Services for IoT processing flow. Amazon Kinesis can deliver IoT data to Amazon cloud by MQTT protocol. To analyze IoT big data delivered by Amazon Kinesis, Amazon Machine Learning provides machine learning functions such as regression or classification.

NTT DOCOMO and GE release an IoT solution which provides GE's industrial wireless router Orbit (MDS-Orbit platform) with NTT DOCOMO's communication module in 2015 [12]. Companies can collect operation statuses of facilities such as bridge, electric and gas by setting Orbit. Moreover, companies can develop IoT applications on Toami [13] which is an IoT cloud platform provided by NTT DOCOMO and enables visualization of collected data easily.

IoT use cases not only analyzing and visualizing IoT data but also taking appropriate actions fast such as automatic repair orders are increased. Therefore, IoT data analysis of stream processing such as Storm [14] or Spark Streaming [15] becomes more popular though batch processing such as MapReduce [16] was major conventionally. Stream processing of IoT data enables fast actions based on real time situation change. HDP [10] is a data processing platform with all Open Source Software stack. HDP provides data analysis modules of batch and stream processing based on HDFS [17] and YARN. Users can analyze data by MapReduce, Spark Streaming or so on, and can store data to HBase, Hive or so on.

Regarding to sensors for acquiring vital data, wearable terminals have been spread. There are various terminals such as watch type, list band type, eyeglass type, T-shirt type and so on. Apple Watch is a watch type computer, contains heartbeat sensor, acceleration sensor and can collect vital data continuously. Sony SmartEyeglass [18] is a eyeglass type computer and can collect acceleration and luminance data. Hitoe [11] is a T-shirt type wearable sensor NTT and Toray develop and can collect Electrocardiograph(ECG) and 3-axis acceleration data by wearing hitoe shirt.

In this way, platforms and sensors have been progressed for vital data analysis. However, when we consider to utilize vital data and take real-time business actions such as substitute member assigns, existing technologies have some problems.

In AWS IoT, to analyze IoT data, users need to collect all data to a cloud and need network cost for many sensors in multiple regions. For example, a satellite communication is used to collect business vehicle's sensing data [9] and network cost is huge. When users analyze collected data, Amazon Machine Learning or Amazon Lambda or other services on a cloud are used, however how to use each service for huge continuous vital data is not considered sufficiently.

IoT applications of the work of [12] developed on Toami are mainly visualize applications of collected data by batch processing. Therefore, applications which take real-time actions such as repair parts orders based on analyzed data are not considered.

Though there are technologies for micro batch or stream processing such as Spark Streaming and Storm, current typical applications are sequential data analysis of SNS posts or operation log analysis. There are few applications for vital data analysis and proper use of micro batch and stream processing to extract necessary data is not discussed sufficiently.

Here, we summarize existing problems. The first is proper use of analyzing methods. To utilize vital data in real-time, not only batch processing but also stream processing for continuous data and micro batch processing for bulk data of short period are needed, but it is not considered to apply health care industry sufficiently. The second is network cost. Because vital data is continuously generated, bandwidth to transfer them to a cloud is large.

## 3. Proposal of vital data analysis platform which resolves existing problems

In this section, we propose vital data analysis platform which resolves existing problems. In 3.1, we explain approaches to resolve existing problems. In 3.2, we show a design of platform to analyze vital data based on HDP.

### 3.1 Approaches to resolve existing problems

Our platform collects workers' vital data via wearable sensor such as hitoe, analyzes them and stores on a cloud. Our platform can store workers' health statuses such as fatigue level change, alerts for emergent situation such as dangerous posture.

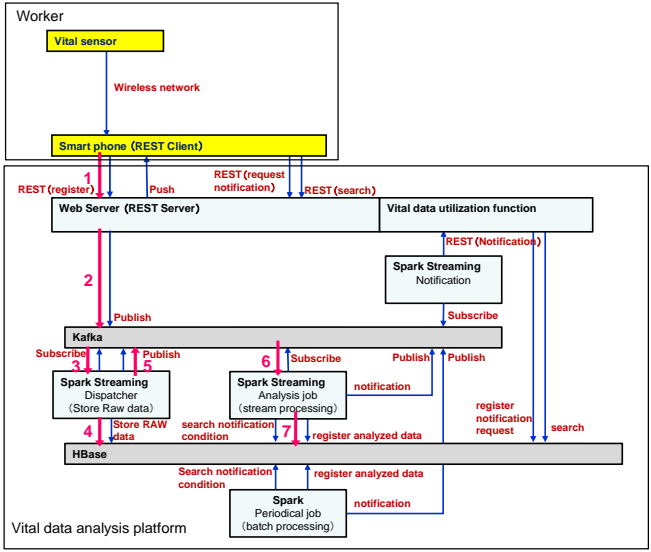

Fig. 1.  Data processing steps based on HDP

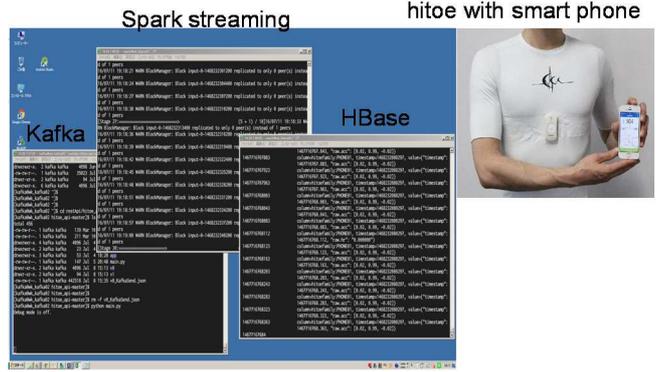

Fig. 2.  Screen image of implemented platform

To resolve existing problems, we propose following two ideas for our platform.

The first is semi real-time analysis of fatigue and relaxation by micro batch processing of vital data using Spark Streaming on a cloud.

The second is stream processing of hitoe data on smart phones to extract primary processed data from raw data and to detect dangerous posture.

Thanks to these ideas, we have following merits.

Smart phones only send primary processed data (e.g., heartbeat interval RRI) from huge continuous data (e.g., ECG) to a cloud. This can reduce network cost. Smart phones also analyze posture in streaming processing and can immediately notify alerts of dangerous postures during working such as pick up things even if a mobile network is disconnected.

A cloud analyzes bulk primary processed data in a short period (e.g., 1 minute RRI data) and extracts high level information such as fatigue or relaxation. Because analyzing fatigue or relaxation needs complex analyzing logic, rich computation resources of cloud are used. And for high accuracy analysis of high level information, analysis of bulk data with certain period is needed. Therefore, we adopt micro batch processing which repeats storing and analyzing in a short period.

### 3.2  Design of vital data analysis platform

Figure 1 shows system image based on above ideas. Figure 1 also shows processing steps of vital sensor data analysis using HDP where sensor stream data is analyzed by Spark Streaming and analyzed data is stored to HBase. Though HDP provides various modules of Open Source Software such as batch processing and SQL processing, we only use modules for target micro batch processing for high level health information. And HDP can be built on cloud middleware such as OpenStack (e.g., OpenStack cloud services such as [19]-[21] and cloud provisioning such as [22][23]).

Wearable sensor hitoe sends workers' vital data to smart phones. Hitoe is an example of wearable sensor which NTT and Toray develop and collects ECG and 3-axis acceleration. Smart phones analyze vital data simply to extract primary processed data of RRI from ECG and posture from acceleration and sends them to a cloud via REST style. This analysis can be done by hitoe SDK [24]. Because raw data of ECG is huge, primary processed data such as RRI is sent to a cloud. If smart phones detect dangerous postures such as picking up things during working, smart phones notify to workers.

Vital data is sent to a cloud and collected data is delivered by messaging system Apache Kafka by publish/subscribe method. Spark Streaming Dispatcher subscribes collected data to Kafka. The Dispatcher stores acquired data to HBase and publishes cleansed data to Kafka. Spark Streaming Analysis job subscribes cleansed data to Kafka. The Analysis job analyzes cleansed data sets of with defined window size in a micro batch processing and extracts fatigue and relaxation level. Window sizes of micro batch are configurable for each extracted data type. Fatigue level is calculated using RRI change (e.g., [25]) and relaxation level is calculated with cardiac vagal index (CVI) [26]. Lastly, analyzed data is stored to HBase.

High level data such as fatigue and relaxation is utilized various ways such as to analyze deeply with other sensor data. When we add other sensor data, we can use cloud configuration technologies of [27] and automatic regression test technologies of [28] for configuration change. And if we need computation power for high level data extraction, we can use GPU to compute such as [29]. And for other systems coordination, Web services [30]-[32] or other technologies such as [33]-[36] can be used.

We implemented our platform using Spark and HBase. Figure 2 shows a screen image of implemented platform and hitoe. We are preparing a field test.

## 4. Related Work

IoT related standardization and industry alliance are too much, and de facto standard is not determined. In application layer, vertical discussions within industry alliance are active such as IIC (Industrial Internet Consortium) and Industrie 4.0 [1]. In platform layer, standardizations are being discussed now such as in OneM2M [37], IEEE P2413 and IEC SG8. Our platform adopts REST or MQTT for sensing data communications because it becomes de facto standard. We need to watch other IoT standardization and de facto standard carefully to improve our platform interoperability.

Amazon, Google and Microsoft also release and provide IoT services. Amazon proposes to apply Amazon service line up to IoT data flow. Kinesis and S3 for collecting and storing data, Lambda for event processing, EMR for data processing and Machine Learning for data analysis. Google releases IoT platform "Brillo" [38]. Brillo expands Android OS for IoT terminals and enables seamless communications between IoT terminals and/or smart phones to create new values. Microsoft provides common IoT functions in Azure IoT services [39] such as notification, stream processing, machine learning and event processing. Based on Azure IoT services, many IoT solutions with partners were achieved. These platforms support real-time processing such as stream data processing but do not consider proper use of stream/micro-batch processing for vital data.

## 5. Conclusions

In this paper, we proposed a vital data analysis platform which resolves existing problems based on HDP architecture to utilize vital data for real-time actions.

Our platform has two characteristics. The first is semi real-time analysis of fatigue and relaxation by micro batch processing of vital data using Spark Streaming on a cloud. The second is stream processing of raw vital data on smart phones. This can reduce network cost to filter huge ECG data and can notify dangerous posture immediately. Our platform can manage workers' health information in real-time with low cost and enables real-time actions.

We will propose our platform to actual industry companies to conduct a field test.

## Acknowledgment

We are thankful Hideki Hayashi who is a manager of this development.

Coordination Service," IEEE International Conference on Web Services (ICWS 2008), pp.600-607, Beijing, Sep. 2008.

(8) H. Sunaga, Y. Yamato, H. Ohnishi, M. Kaneko, M. Iio and M. Hirano, "Service Delivery Platform Architecture for the Next-Generation Network," ICIN2008, Session 9-A, Bordeaux, Oct. 2008.

(9) S. Arakawa, "Development and Deployment of KOMTRAX Step 2," Komatsu Technical Report Vol.48, No.150, 2002.

(10) Hortonworks data platform website, http://hortonworks.com/

(11) S. Tsukada, H. Nakashima and K. Torimitsu, "Conductive Polymer Combined Silk Fiber Bundle for Bioelectrical Signal Recording", PLoS ONE, Vol.7, No.4, pp.e33689, 2012

(12) NTT Docomo press release website, https://www.nttdocomo.co.jp/english/info/media_center/pr/2015/0708_00.html

(13) Toami web site (Japanese), http://www.m2m-cloud.jp/

(14) N. Marz. "STORM: Distributed and fault-tolerant realtime computation," 2013. http://storm-project.net.

(15) M. Zaharia, M. Chowdhury, M. J. Franklin, S. Shenker and I. Stoica, "Spark: Cluster computing with working sets," Proceedings of the 2nd USENIX Conference on Hot Topics in Cloud Computing, 2010.

(16) J.Dean, and S. Ghemawat, "MapReduce: Simplified data processing on large clusters," In proceedings of the 6th Symposium on Opearting Systems Design and Implementation (OSDI'04), pp.137-150, Dec. 2004.

(17) K. Shvachko, Hairong Kuang, S. Radia and R. Chansler, "The Hadoop Distributed File System," IEEE 26th Symposium on Mass Storage Systems and Technologies (MSST2010), pp.1-10, May 2010.

(18) Sony SmartEyeglass web site, http://developer.sonymobile.com/products/smarteyeglass/

(19) Y. Yamato, Y. Nishizawa, S. Nagao and K. Sato, "Fast and Reliable Restoration Method of Virtual Resources on OpenStack," IEEE Transactions on Cloud Computing, DOI: 10.1109/TCC.2015.2481392, 12 pages, Sep. 2015.

(20) Y. Yamato, N. Shigematsu and N. Miura, "Evaluation of Agile Software Development Method for Carrier Cloud Service Platform Development," IEICE Transactions on Information & Systems, Vol.E97-D, No.11, pp.2959-2962, Nov. 2014.

(21) Y. Yamato, S. Katsuragi, S. Nagao and N. Miura, "Software Maintenance Evaluation of Agile Software Development Method Based on OpenStack," IEICE Transactions on Information & Systems, Vol.E98-D, No.7, pp.1377-1380, July 2015.

(22) Y. Yamato, Y. Nishizawa, M. Muroi and K. Tanaka, "Development of Resource Management Server for Production IaaS Services Based on OpenStack," Journal of Information Processing, Vol.23, No.1, pp.58-66, Jan. 2015.

(23) Y. Yamato, "Performance-Aware Server Architecture Recommendation and Automatic Performance Verification Technology on IaaS Cloud," Service Oriented Computing and Applications, Springer, DOI: 10.1007/s11761-016-0201-x, Nov. 2016.

(24) hitoe transmitter SDK web site, https://dev.smt.docomo.ne.jp/?p=docs.api.page&api_name=iot_control&p_name=api_usage_scenario

(25) K. Yokoyama and I. Takahashi, "Feasibility Study on Estimating Subjective Fatigue from Heart Rate Time Series," IEICE Transactions on Fundamentals of Electronics, Communications and Computer Sciences, Vol.96, No.11, pp.756-762, 2013. (in Japanese)

(26) Stephen. W. Porges, "Cardiac vagal tone: a physiological index of stress," Neuroscience and Biobehavioral Reviews, Vol.19, No.2, pp.225-233, 1995.

(27) Y. Yamato, M. Muroi, K. Tanaka and M. Uchimura, "Development of Template Management Technology for Easy Deployment of Virtual Resources on OpenStack," Journal of Cloud Computing, Springer, 2014, 3:7, DOI: 10.1186/s13677-014-0007-3, 12 pages, June 2014.

(28) Y. Yamato, "Automatic verification technology of software patches for user virtual environments on IaaS cloud," Journal of Cloud Computing, Springer, 2015, 4:4, DOI: 10.1186/s13677-015-0028-6, 14 pages, Feb. 2015.

(29) Y. Yamato, "Optimum Application Deployment Technology for Heterogeneous IaaS Cloud," Journal of Information Processing, Vol.25, No.1, Jan. 2017.

(30) Y. Nakano, Y. Yamato, M. Takemoto and H. Sunaga, "Method of creating web services from web applications," IEEE International Conference on Service-Oriented Computing and Applications (SOCA 2007), pp.65-71, Newport Beach, June 2007.

(31) Y. Yamato, H. Ohnishi and H. Sunaga, "Study of Service Processing Agent for Context-Aware Service Coordination," IEEE International Conference on